\documentclass[12pt,a4paper,twoside]{paper}
\usepackage{amsmath,bbm,amssymb,amsxtra}
\usepackage{graphicx}
\usepackage{epstopdf}
\usepackage{chngcntr}
\usepackage{etoolbox}

\usepackage{bbold}

\usepackage{fancyhdr}
 \usepackage{fancyhdr}
\newcommand{\be}{\begin{equation}}
\newcommand{\ee}{\end{equation}}

\newcommand{\ba}{\begin{eqnarray}}
\newcommand{\ea}{\end{eqnarray}}

\newcommand{\bea}{\begin{eqnarray*}}
\newcommand{\eea}{\end{eqnarray*}}

\newcommand{\bee}{\begin{enumerate}}
\newcommand{\ene}{\end{enumerate}}

\usepackage{epsfig}
\def\R{\mathbb R}

\usepackage{appendix}
\usepackage{chngcntr}
\usepackage{etoolbox}
\usepackage{lipsum}
\usepackage{amsmath,bbm,amssymb,amsxtra}
\usepackage{enumerate}
\allowdisplaybreaks
\usepackage[active]{srcltx}
\usepackage{epsfig}
\usepackage{ae}
\numberwithin{equation}{section}

\newtheorem{lemma}{Lemma}[section]
\newtheorem{proposition}[lemma]{Proposition}

\newtheorem{theorem}[lemma]{Theorem}

\newtheorem{remark}[lemma]{Remark}
\newtheorem{definition}[lemma]{Definition}

\newcommand{\prop}[1]{\begin{proposition}\label{#1}
\sl }
\newcommand{\eprop}{\end{proposition}}
\newcommand{\thm}[1]{\begin{theorem}\label{#1}
\Ä }
\newcommand{\ethm}{\end{theorem}}
\newcommand{\lem}[1]{\begin{lemma}\label{#1}
\sl }
\newcommand{\elem}{\end{lemma}}

\newcommand{\defin}[1]{\begin{definition}\label{#1}
\sl }
\newcommand{\edefin}{\end{definition}}

\def\la{\lambda}

\def\CA{{\mathcal A}}

\def\qq{ \begin{eqnarray} }
\def\qqq{ \end{eqnarray} }
\def\rr{ \begin{equation} }
\def\rrr{ \end{equation} }

\def\qq{ \begin{eqnarray} }
\def\qqq{ \end{eqnarray} }

\usepackage{epsfig}

\counterwithin*{equation}{section} 
\counterwithin*{equation}{subsection} 

\begin{document}

    \begin{center}
 \textsl{\huge    The De Broglie-Bohm theory {\it is} and {\it is not}  a hidden variable theory.\footnote{ This is a draft of a chapter that has been accepted for publication by Oxford
University Press in the forthcoming book ``Guiding Wave in Quantum Mechanics: 100 Years of de Broglie-Bohm Pilot-Wave Theory" edited by Andrea Oldofredi,  due for
publication in 2023.}}

 \vspace*{8mm}
 { \Huge Jean Bricmont\footnote{IRMP,
Universit\'e catholique de Louvain,
Chemin du Cyclotron 2,
1348 Louvain-la-Neuve,
Belgium. E-mail: jean.bricmont@uclouvain.be},

}
\end{center}

\begin{abstract}

 We will first define what is meant by ``hidden variables". Then, we will review various theorems proving the impossibility of theories introducing such variables and then show that the de Broglie-Bohm theory is not refuted by those theorems. We will also explain the relation between those theorems and nonlocality, with or without introducing Bell's inequalities.
 
\end{abstract}

\section{Introduction}\label{sec1}

The title of this article is obviously self-contradictory; but, as an American president might have said, what do you {\it mean} by ``hidden variable theory"? One of the goals of this paper will be to clarify this issue and to try to dispel the many confusions surrounding this expression.

 For example, in the wake of the awarding of the 2022 Nobel Prize in Physics to John Clauser, Alain Aspect and Anton Zeilinger, for verifying the violation of Bell's inequalities, it has  frequently been asserted that those violations imply the impossibility of hidden variables theories (that could be alternatives to ordinary quantum mechanics)\footnote{For example, in its press release announcing the 2022 Nobel Prizes in physics, the Royal Swedish Academy writes: ``This means that quantum mechanics cannot be replaced by a theory that uses hidden variables.", see https://www.nobelprize.org/prizes/physics/2022/press-release/.}, or maybe only to local hidden variables theories. The latter qualification is often associated to the idea that nonlocal hidden variables theories are too weird to consider.

Yet it is known since 1952, through the work of David Bohm, and in fact since 1927, through the work of Louis de Broglie, that a ``hidden variable" alternative to ordinary quantum mechanics exists. But why isn't it excluded by the violation of Bell's inequalities? Is it because it is nonlocal? But, if yes, is that a defect or a quality of the de Broglie-Bohm theory?

\section{What are ``hidden variables"? }\label{sec2}

By definition, the expression ``hidden variables" refers to any variable that would describe the physical state of a system beyond its quantum state.

But a moment's reflection shows that, if we accept this definition, the world is full of hidden variables. To illustrate this idea, consider the famous example of Schr\"odinger's cat. Since this example is well known, let us simply recall that, at the end of an experiment where one measures the property of a particle that can take two values, the measuring device, or the cat if we couple the device to the cat through a poison capsule, is (leaving aside normalization factors):
 
\be
\Psi_{\mbox{cat alive}}+\Psi_{\mbox{cat dead}}.
\label{1}
\ee
And that, argued Schr\"odinger, cannot be a complete description of the cat, which is obviously either alive or dead but not both! The way out of this problem from the point of view of ordinary quantum mechanics is to introduce the collapse postulate: when one looks at the cat, one sees whether she is alive or dead and, depending on what one sees, one reduces the wave function of the cat (and of the particle that was measured and is thus coupled to the state of the cat) to either $\Psi_{\mbox{cat alive}}$ {\it or} $\Psi_{\mbox{cat dead}}$.

Since this is a {\it deus ex machina} from the point of view of the linear Schr\"odinger evolution, justifying it is often viewed as the main problem in foundations of quantum mechanics.

Saying that the cat is alive or dead (depending on what we see) amounts to the introduction of a ``hidden variable" since the property of being dead or alive is not part of the quantum state (\ref{1}).

But actually there is a deeper problem: neither $\Psi_{\mbox{cat alive}}$ nor $\Psi_{\mbox{cat dead}}$ are cats: they are functions defined on a high dimensional space $\R^{3N}$, $N$ being the number of particles composing the cat (putting aside the fact that there are different kinds of particles and that  they may also have a spin) while cats are located in $\R^3$. And it is not clear what it means to say that $\Psi_{\mbox{cat alive}}$ or $\Psi_{\mbox{cat dead}}$ are descriptions of cats, let alone ``complete descriptions" of them\footnote{This idea is emphasized by Tumulka \cite[Sect. 5.1]{Tu2}}. Like all wave functions they allow us to predict results of measurements done on the objects that they ``describe", but nothing else.

What most people do is to mentally identify cats and wave functions of cats. But the cat, simply by being situated in 
the usual space $\R^3$, has properties that are not part of the wave functions $\Psi_{\mbox{cat alive}}$ or $\Psi_{\mbox{cat dead}}$.

The cat has a shape, a size, a weight etc. that are, according to our definition, hidden variables. This does not mean that one could not, in principle, calculate those properties from a detailed knowledge of the wave function of the cat, but that these properties refer to objects in three dimensions and not to functions defined on $\R^{3N}$.

In the very useful terminology of John Bell, the cat (or any other macroscopic object) is a {\it local beable}. The word beable was created in opposition to the word observable, to refer to what exists in the world, and not simply to what is observed (by us, humans) and the world local refers to the fact that those beables are situated in the ordinary space $\R^3$.

\section{Na\"ive Statistical Interpretation}\label{sec3}
  
If we admit that there exist ``hidden variables" in the macroscopic world (variables that, in fact, exist everywhere, and are not hidden at all but are the only things we ever see), why not try to assume that they exist also in the microscopic world?
  
This leads us to the most natural interpretation of quantum mechanics, which is probably the one in the mind of most of the ``no worry about quantum mechanics" physicists (and also  the one of Einstein), namely the ``statistical" one:

According to that interpretation, a quantum state $\Psi$ does not represent an individual system but an {\it ensemble} of systems.
 
And, for any ``observable" represented by an operator $A$, there is a well-defined value $v(A)$ that a measurement of $A$ for an individual quantum system would reveal (and not create, since it  pre-exists to any measurement).
 That value $v(A)$ is again  a ``hidden variable", because it is a property of the system not included in the quantum state. In other words, there would be a ``hidden variable" $v(A)$ for each operator $A$.

And we assume, in this interpretation, that, if 

\be
\Psi= \sum_n c_n \Psi_n,
\label{2}
\ee
where the $\Psi_n$'s are the eigenvectors of $A$, with eigenvalues $\la_n$, then the frequency with which one obtains the result $v(A)=\la_n$ in an ensemble defined by $\Psi$ is $|c_n|^2$, so that Born's rule holds.

In that view, there is no problem with the collapse or reduction of the quantum state: one updates one's probabilities given some new information; if we learn that the value associated to the physical quantity represented by the operator $A$ is $v(A)=\la_n$, then we update our quantum state so that it becomes the corresponding eigenvector $\Psi_n$.

All would be well if this interpretation was tenable; unfortunately, it is not, for a purely mathematical reason.
 Indeed, it is a well-known property of quantum mechanics that, if the  operators 
$A$ and $B$
 commute ($ [A, B]= AB-BA=0$),  then, they are simultaneously measurable, and  the values $v(A)$, $v(B)$, etc. revealed by those measurements would have to satisfy
 \be
v(AB)=v(A) v(B),
\label{3}
\ee
and
 
 \be
v(A+B)=v(A)+ v(B).
\label{4}
\ee
 
However, the following theorem renders the na\"ive statistical interpretation untenable.

\begin{theorem}(No hidden variables theorem)\label{t1}
	
Let $\cal A$ be the set of self-adjoint operators on a  Hilbert space $\cal H$  with $\dim \cal H$ at least equal to 4.

Then,  there does not  exist a map  $v: \cal A \to \R$ such that:

\begin{itemize}
\item[1.] $ \; \forall A \in { \cal A}$, 
\be
v(A) \;\;\mbox{is an eigenvalue of} \;\;A.
\nonumber
\ee

\item[2.] $\forall A, B \in { \cal A}$ with  $ [A, B]= AB-BA=0$, (\ref{3}) or (\ref{4}) are satisfied.
\end{itemize}

\end{theorem}

\begin{remark}
 Theorem \ref{t1} is due to John Bell \cite{Be1} and to Kochen and Specker \cite{KS}. For a simple proof due to David Mermin, see \cite{Me4}.
 
To prove the theorem, we do not need to assume that the map is defined on all operators in $\CA$. For example, the Kochen and Specker proof \cite{KS} uses the squares of spin matrices, $S_x, S_y, S_z$, for spin associated to any  three dimensional set of orthogonal vectors $x, y, z$ in $\R^3$. They have the following properties:
\begin{itemize}
\item[1.]The eigenvalues of $S_x^2$, $S_y^2$ and $S_z^2$ are  $0$  and $1$.
\item[2.] 
$[S_x^2, S_y^2]= [S_y^2, S_z^2]= [S_z^2, S_x^2]= 0.$
\item[3.] 
$
S_x^2+S_y^2+S_z^2=2.$
\end{itemize}

From that and assumptions 1 above and (\ref{4}), it follows that the triple $(v(S_x^2), v(S_y^2), v(S_z^2))$ must be either $(1, 1, 0)$ or $(1, 0, 1)$ or $(0, 1, 1)$.

But that must hold {\it for every}  set of  three dimensional orthogonal vectors $x, y, z$ in $\R^3$. Kochen and Specker were able to exhibit a finite number of such sets so that the above assumption on the values taken by $(v(S_x^2), v(S_y^2), v(S_z^2))$ leads to a contradiction\footnote{In their original argument, Kochen and Specker used 117 such sets \cite{KS}, but that number was reduced to 33 by Peres \cite{Per, Per1}. Note in passing that the Kochen-Specker argument only requires the Hilbert space $\cal H$ to have  a dimension  at least equal to $3$. But the proof of Mermin \cite{Me4} that requires a Hilbert space of dimension  at least equal to $4$ is simpler.}.

\end{remark} 

The obvious conclusion of this theorem is that one cannot have a statistical distribution of maps that do not exist!

A more general conclusion is that one cannot assume that measuring devices really measure properties of physical systems, that pre-exist to their ``measurement". 

Measuring devices must have an active role. This was actually an intuition of Bohr:

 \begin{quote}

[\dots ] the {\it impossibility of any sharp distinction between the behavior of atomic objects and the interaction with the measuring instruments which serve to define the conditions under which the phenomena appear}.

\begin{flushright}  Niels Bohr  \cite[p.~210]{Bohr}, quoted in \cite[p.~2]{B} (italics in the original)\end{flushright}
\end{quote}

But how do these devices interact with the physical system? In ordinary quantum mechanics there is no answer to that question, because the ``measurements" lead to collapses of the quantum state, which is a {\it deux ex machina} from the point of view of the Schr\"odinger evolution. We need a more detailed theory that explains what happens during ``measurements".  

Let us now turn to a statistical and ``hidden variables" theory that provides exactly such an explanation.

\section{The de Broglie-Bohm theory}\label{sec4}

The theory of de Broglie (1927) (see his contribution to the 1927
 Solvay Conference in \cite{BV}), and Bohm (1952)\cite{Bo1}, popularized by Bell \cite{B}, and by D\"urr, Goldstein and Zangh\`i \cite{DGZ1}\footnote{There are many expositions of the de Broglie--Bohm theory, see, e.g., \cite{Al, Tu} for elementary introductions and \cite{BV, BH, Bri1, DGZ, DT, DL, Ho, Go, Ma1, Nor, Tu1} for more advanced ones. There are also pedagogical videos made by students in Munich, available at: \\ \texttt{https://cast.itunes.uni-muenchen.de/vod/playlists/URqb5J7RBr.html}.}:

\begin{itemize}
\item[1.]
Is a theory of ``hidden variables" (although they are not at all hidden), 

\item[2.]
That accounts for all the phenomena predicted by ordinary
quantum mechanics, 

\item[3.]
That is not contradicted by the no hidden variables theorems,

 \item[4.]
And that explains why measurements do not in general measure pre-existing properties of a system (in other words, it explains why measuring devices have an ``active role").
 \end{itemize}
 
To start, let us think of the double slit experiment, 
see Fig.~\ref{fig1}:

\begin{figure}[ht]
\centering
\includegraphics[keepaspectratio,height=10cm]{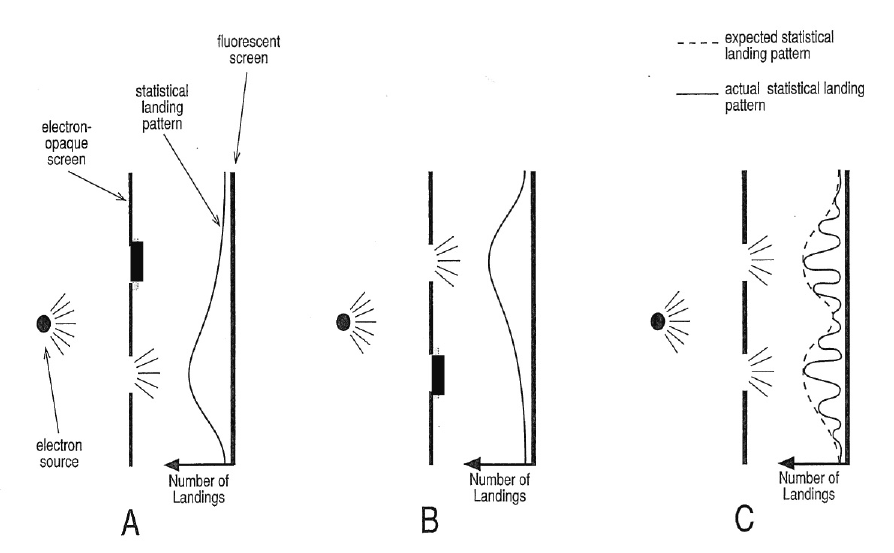}\caption[]{A schematic description of the distribution of particles on the second wall on the right of the figure,  depending on whether one or two slits are open in the first wall.}\label{fig1}
\end{figure}

How can electrons be both particles and waves?
The answer of de Broglie-Bohm is that they are particles {\it guided} by waves.

In the de Broglie-Bohm's theory, the state of system is  a pair $(X, \Psi)$, where $X=(X_1,\dots,X_N)$ denotes the actual positions of all the particles in the system under consideration, that exist, whether we measure them or we ``look" at them or not.

Here, 
$\Psi= \Psi (x_1,\dots, x_N)$ is the usual wave function, $(x_1,\dots, x_N)$ denoting the arguments of the function   $\Psi$. $X$ are the ``hidden variables" in this theory; this is obviously a misnomer, since particle positions are the only things that we ever directly observe (think of the double-slit experiment for example).

 	The dynamics of the de Broglie-Bohm's theory is as follows\footnote{For simplicity, we limit ourselves to the non-relativistic case and without spin. Extensions to relativistic equations exist (see e.g. Tumulka \cite[Sect. 7.6]{Tu1}) and we will introduce the spin in Sect. \ref{sec5}, but without equations.};
both objects  $\Psi$ and $X$ evolve in time: 

\begin{itemize}
	\item [1.]
The wave function, at all times, and whether one measures something or not
evolves according to Schr\"odinger's equation.

The wave function never collapses for a closed system\footnote{However, in situations that correspond to measurements in ordinary quantum mechanics, there is an effective collapse of the wave function, see D\"urr et al. \cite[Section 5]{DGZ}.}.

\item [2.]  The evolution of the positions is  guided by the wave function: writing $\Psi= R e^{iS}$:
\be
\frac{d}{dt}  X_k(t) =
 \nabla_k S (X_1(t),\ldots,X_N (t))
 \label{5}
\ee
for $k=1, \ldots, N$, where $X_1(t),\ldots,X_N(t)$ are the actual positions of the particles at time $t$.  
\end{itemize}

\begin{figure}[ht!]
\centering
\includegraphics[keepaspectratio,height=10cm]{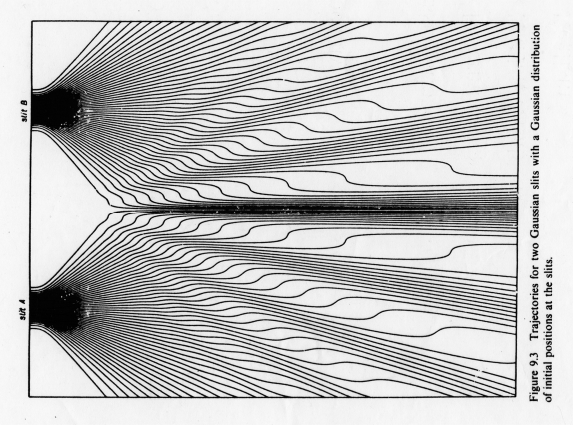}
\caption[]{De Broglie--Bohm trajectories computed for the double-slit experiment. Each line represents a particle trajectory.  Reproduced with the kind permission of {\it Societ\`a Italiana di Fisica\/} and the authors, from \cite{PDH}}\label{fig2}
\end{figure}

A classical way to illustrate the dynamics of that theory is given by the  numerical solution of the double slit experiment shown in Fig. \ref{fig2}:

The motion {\it in vacuum}, behind the slits, is highly {\it non classical}, i.e. not rectilinear! 
Note  that one can determine  a posteriori through which hole the  particle went. Note also the presence of a nodal line: it turns out that, by  symmetry of  $\Psi$, the velocity is tangent to the middle line; thus, particles cannot cross it.
 
This situation was nicely described by John Bell:

 \begin{quote}
Is it not clear from the smallness of the scintillation on the screen
that we have to do with a particle? And
is it not clear, from the diffraction and interference patterns, that the
motion of the particle is directed by
a wave? De Broglie showed in detail how the motion of a particle, passing
through just one of two holes in
the screen, could be influenced by waves propagating through both holes.

And so
influenced that the particle does
not go where the waves cancel out, but is attracted to where they
cooperate. This idea seems to me so natural
and simple, to resolve the wave-particle dilemma in such a clear and
ordinary way, that it is a great mystery to
me that it was so generally ignored.  

\begin{flushright} J. Bell \cite[p. 191]{B}\end{flushright}

\end{quote}

A natural question is: how does the theory of de Broglie-Bohm account for the statistical predictions of quantum mechanics?

That is due to a property of the de Broglie-Bohm dynamics called {\it equivariance}, which is illustrated in Fig. \ref{fig3}:

\begin{figure}[t]
\centering
\includegraphics[keepaspectratio,height=10cm]{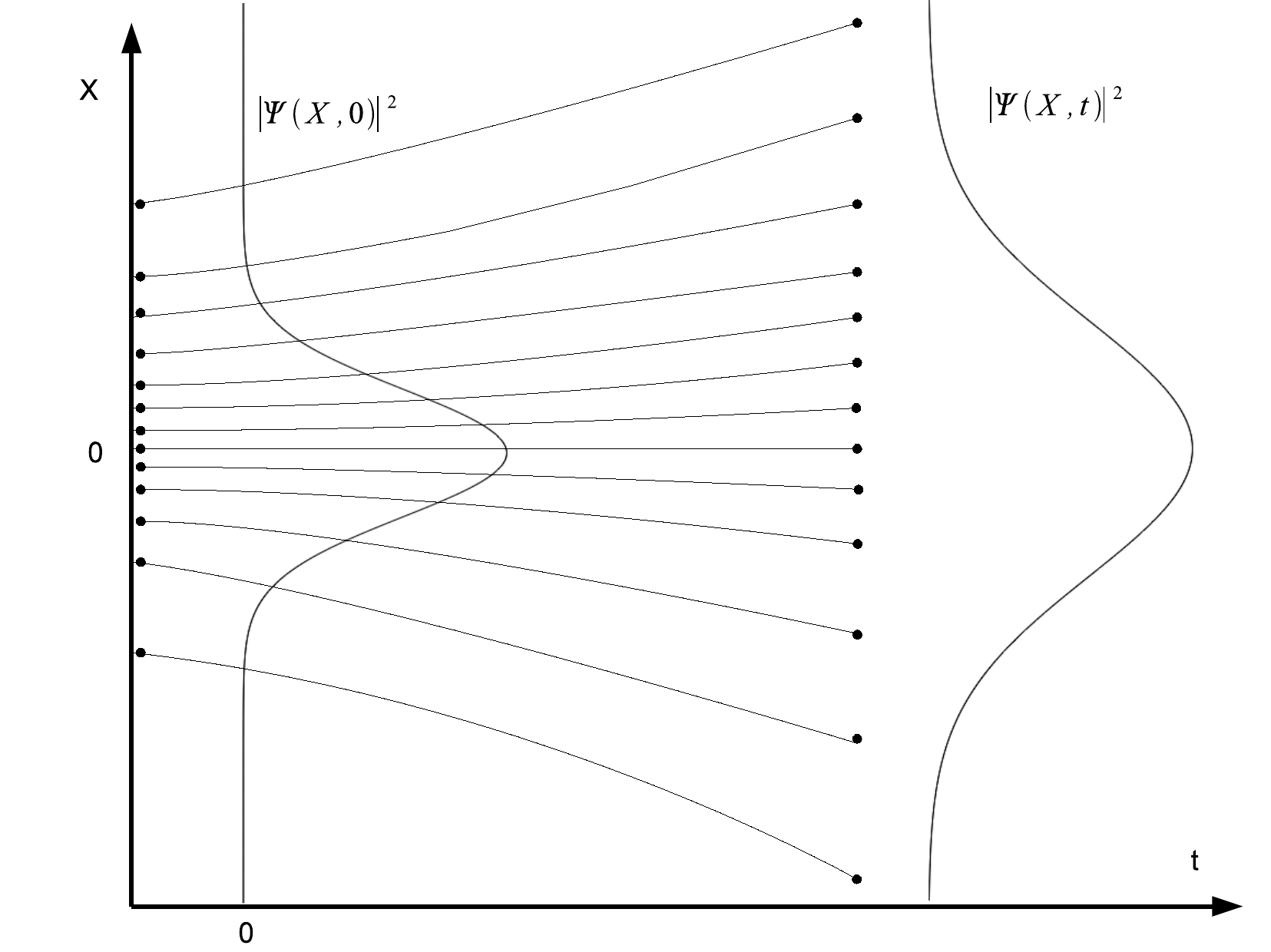}\caption[]{Illustration of the property of equivariance of the $|\Psi ( X, t)|^2$ distribution, in one dimension, for a Gaussian $\Psi$. Each dot represents the position of a particle, both at time $0$ and at time $t$, connected by trajectories.}\label{fig3}
\end{figure}

Assume that the initial density of particles $\rho_{0}$ is (approximately) given by  $\rho_{0}(X)= |\Psi (X, 0)|^2$, see  the left of Fig. \ref{fig3}.
Then, the empirical density of particles at later times
 $\rho_{t}$ will satisfy $\rho_{t}(X)= |\Psi (X, t)|^2$,
 where $\Psi (X, t)$ is the solution of the Schr\"odinger equation and 
 $\rho_{t}$  comes 
  from the guiding equation \ref{6}.
This is illustrated on the right of Fig. \ref{fig3}.

So, if we assume that $\rho_0= |\psi_0|^2$ at some initial time, it will hold at all times. 
The statistical predictions of quantum mechanics are recovered in the de Broglie-Bohm theory, at least as far as the positions of the particles are concerned. But they are also recovered for all other observables, through an analysis of ``measurements" in that theory, as we will see in the next section.

The assumption that $\rho_0= |\psi_0|^2$ is called quantum equilibrium. In any deterministic dynamical system, statistics of results depend on statistical assumptions on the initial conditions of the system (think of statistical mechanics as an example).

And, ultimately, any assumptions on initial conditions, bring us, through a version of the egg and the hen problem, to assumptions on the initial conditions of the universe, which is a complicated issue that we will not discuss here (see D\"urr et al. \cite{DGZ}). But, from a somewhat pragmatic point of view, one can put aside the egg and the hen problem, and simply assume a quantum equilibrium distribution of the particles' positions at the beginning of any physical situation that one analyzes.

A natural question, to which we will now turn is: why isn't the de Broglie-Bohm theory refuted by the no hidden variables theorems, since it is a hidden variables theory?

\section{The de Broglie-Bohm theory and the no hidden variables theorems}\label{sec5}

The answer to that question will reveal the main quality  of de Broglie-Bohm's theory, namely that it explains what happens during ``measurements".

To illustrate this, wee will consider two examples: 

First, a ``measurement" of something that does not exist: the spin.  Then, a ``measurement" of something that does exist but is not measured by that ``measurement": the momentum.

\begin{figure}[ht]
\centering
\includegraphics[width=0.9\textwidth]{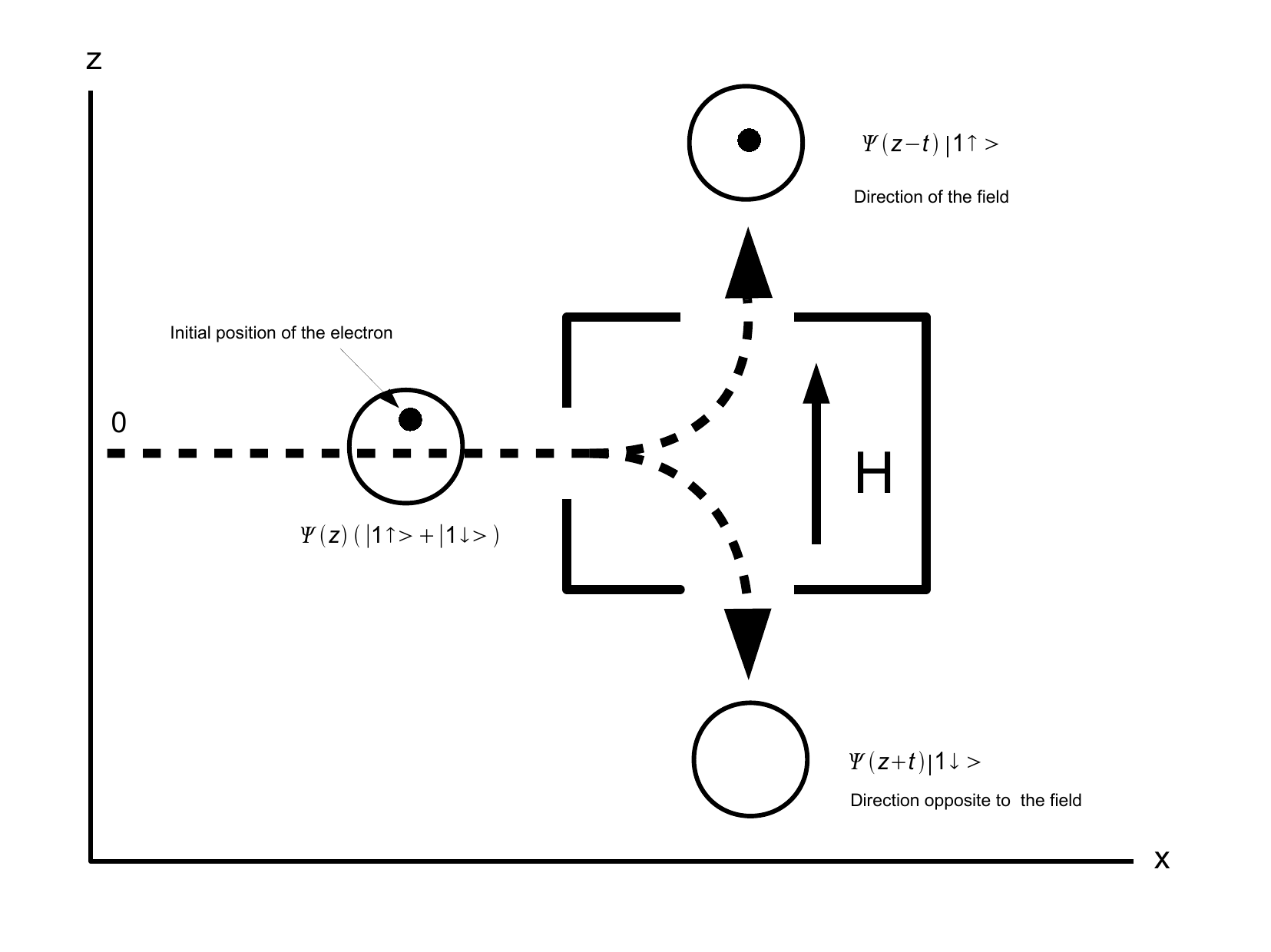}\caption[]{An idealized spin measurement.}\label{fig4}
\end{figure}

 To give an example of a ``measurement" of something that does not exist\footnote{This example is taken from Albert \cite[p. 153]{Al}; see also Daumer et al. \cite{NRAO}.}, consider  a Stern-Gerlach apparatus ``measuring" the spin\footnote{In Sect. \ref{sec4}, we did not write down the equations for particles with spin, see e.g. D\"urr and Teufel \cite[Section 8.4]{DT} for the de Broglie-Bohm theory of particles with spin.}, see Fig. \ref{fig4}, where $H$ be the magnetic field.

 The $|1 \uparrow>$ part of the state always goes in  the direction of the field,
 and the $|1 \downarrow>$ part always goes in the opposite direction.

 If the  particle is initially in the upper part of the support of the wave function (for a  symmetric wave function), it will always go upwards. That is just a property of the de Broglie-Bohm dynamics.

\begin{figure}[ht]
\centering
\includegraphics[width=0.9\textwidth]{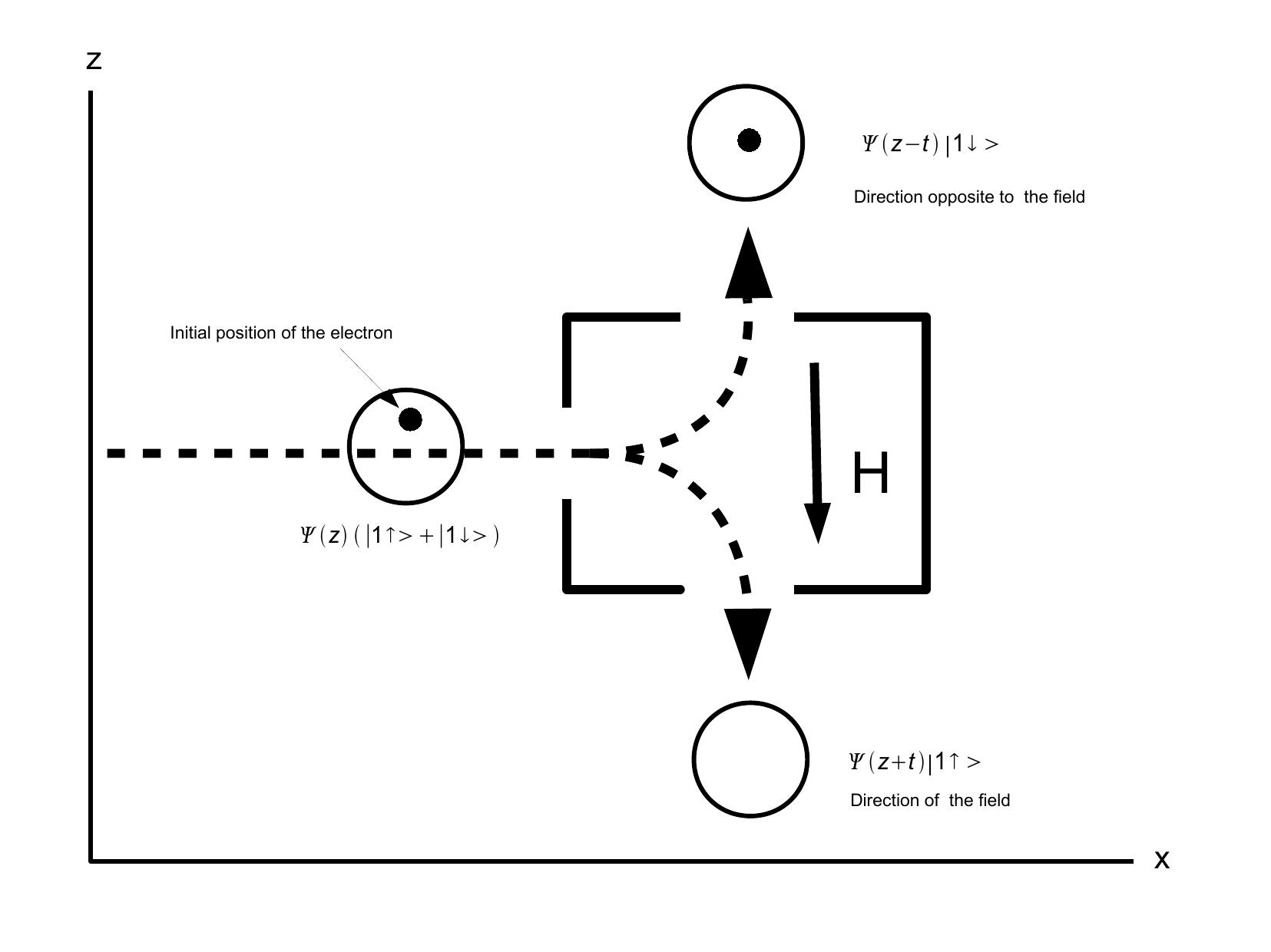}[\caption[]{An idealized spin measurement, with the orientation of the field reversed compared to Fig. \ref{fig4}.}\label{fig5}

\end{figure}
 
Now, repeat the same experiment, but with  the direction of the gradient of the field reversed (see Fig. \ref{fig5}), and let us assume that the particle starts with {\it exactly the same wave function and the same position as before}.

If the  particle is initially in the upper part of the support of the wave function, it will again go upwards.
That is because there is here, as in Fig. \ref{fig2}, a nodal line in the middle of the figure that particles cannot cross.
 But going upwards means now going in the direction {\it opposite} to the gradient of the field (since the latter is reversed).

 So, the particle whose spin was ``up", will  ``have" its spin ``down", although one ``measures" exactly the same quantity (the spin in the vertical direction), with {\it  exactly the same initial conditions (for both the  wave function and the  position of the particle)}, but with
two different arrangements of the apparatus. So, by changing only that arrangement, we get different  results, for the same initial conditions of the particle.

 This is simple illustration of the active role of the measuring device.
Another way to say this: {\it ``spin is not real" }. Of course, the what is real is that the quantum state has a spinor part, but what is not real is any pre-existing value of the spin.
 
 Or, more precisely:
{\it  quantum ``measurements" are interactions that do not measure something intrinsic to the particle}.

To give now an example of a ``measurement" of something that does exist but is not measured by that ``measurement", consider a particle in a box.

 If the wave function is real (for example, the ground state), then the particle is at rest\footnote{For spinless particles. If one includes the spin, particles are not at rest, see Das \cite{Das}.}. Indeed, write (for one particle):
$\Psi= R e^{iS}$,  and
\be
\frac{d}{dt}  X(t) =
 \nabla S (X(t)).
 \label{6}
\ee
If $S=0$, $\frac{d}{dt}  X(t)=0$. 

So, we know its velocity: it is zero!
Note that we can also, if we want, measure its position to arbitrary accuracy.
It looks like this violates Heisenberg's inequality! 
But deducing the velocity from the theory is not what ``measuring the velocity" means in quantum mechanics.

To do that ``measurement", one needs to open the box, let the particle move freely; measure its position at some later time and divide the distance travelled by the elapsed amount of time.

This gives us the ``measured velocity", which is not zero and whose statistical distribution satisfies Heisenberg's inequality, but that has nothing to do with the ``true" velocity of the particle before the ``measurement", which, as we saw, was zero\footnote{For the calculation of this result, see e.g. \cite[Section 5.D.2]{Bri1}.}. 
 
 All this vindicates the idea that   ``measurements" are genuine interactions between a system and an apparatus, which was actually an intuition of Bohr mentioned at the end of Section \ref{sec3}. But now, this follows from the equations of the theory and not from some  philosophical {\it a priori}.

John Bell emphasized also the active role of measurements:

\begin{quote}
A second charge [against the word measurement] is that the word comes loaded with meaning from everyday life, meaning which is entirely inappropriate in the quantum context. When it is said that something is 'measured' it is difficult not to think of the result as referring to some pre-existing property of the object in question. This is to disregard Bohr's insistence that in quantum phenomena the apparatus as well as the system is essentially involved.

\begin{flushright}  John Bell  \cite{Be2}\end{flushright} 
\end{quote}    
  
Since ``measurements" are interactions that do not reveal pre-existing properties of the system, the de Broglie-Bohm's theory is not refuted by the no hidden variables theorems (see also Daumer et al. \cite{NRAO} for a discussion of this fact).

The de Broglie-Bohm's theory is a statistical theory (the distribution of the particles' positions is ``random"), but a consistent one, because it does not associate pre-existing values to ``observables" other than positions.

One can (rather easily) show that, if one assumes quantum equilibrium for the distribution of the particles' positions, then Born's rule is satisfied for the ``measurements" of all ``observables", see D\"urr et al. \cite{DGZ, DG4} for a general proof of this statement.

\section{The no hidden variables theorems and nonlocality }\label{sec6}

A standard objection against the de Broglie-Bohm's theory is that it is nonlocal: indeed, in  (\ref{5}), the velocity of a given particle at time $t$ depends on the positions of all the other particles of the system, at that same time $t$, no matter how far apart the particles may be. So that, if one acts on a given particle so as to change its position, that affects instantaneously the velocities  of all the other particles in the system. For a concrete illustration of how this nonlocality expresses itself, see \cite[Section 5.2.1]{Bri1}.

We will argue now that this nonlocal character is a quality rather than a defect of the de Broglie-Bohm's theory, because Nature itself is nonlocal, as shown by a combination of an argument due to Einstein, Podolsky and Rosen (EPR) \cite{EPR} and another one due to Bell \cite{Be3, GNTZ}. We will also see how to simplify Bell's argument using the no hidden variables theorems. We start with the EPR argument (in a slightly modified form and using the Bohm version of their argument \cite{Bo}, instead of the original one).

\subsection{The EPR dilemma}\label{sec6.1}

Consider a pair of particles, one of which is sent to Alice and another to Bob, (who are situated far apart), and  whose quantum state is an entangled one:
 
 \begin{eqnarray}
|\Psi\rangle &=& \frac{1}{\sqrt 2} \big(| A \uparrow  \rangle| B \downarrow  \rangle-| A \downarrow  \rangle|  B \uparrow  \rangle\big),
\label{6a}
\end{eqnarray}
with $A$ being the particle sent to Alice and $B$  the one sent to Bob. That state, according to ordinary quantum mechanics, means that the spin measured by Alice, in any given direction,  will have equal probability to be up or down, but is perfectly correlated with the one of Bob: if the spin is up for Alice (the state $| A \uparrow  \rangle$), it will be down for Bob  (the state $| B \downarrow  \rangle$) and vice-versa.

Now comes the crucial question: how does one account for these perfect correlations?

There seems to be only two possibilities:
\begin{itemize}
	\item [1.]

Either each particle carries with it instructions according to which, upon measurement, its spin will be up or down and the instructions of both particles are correlated.

\item [2.] Or, the measurement of the spin on Alice's side influences instantaneously the result on Bob's side (or vice-versa), and this influence guarantees that the result of  Bob's spin measurement is  correlated with the one of Alice.

\end{itemize}

This is what we  call {\bf EPR's Dilemma}.
  But EPR did not consider it as a dilemma, because they only considered the first possibility as acceptable, since the second possibility amounts to admitting the existence of instantaneous actions at a distance that EPR could not even envision, since the existence of those actions seemed to contradict the theory of relativity.

The instructions mentioned in the first possibility are called ``hidden variables" and assuming their existence means that the description of the physical situation by the quantum state is incomplete (which is what EPR thought to have proven).

To be historically correct, this is EPR's (logical) conclusion (without speaking of a dilemma), but not quite their argument\footnote{Moreover, this is Bohm's version of their argument; EPR considered  positions and momenta of correlated particles instead of the spin variables.}, which was more complicated and unnecessarily so. 
 
 \subsection{What did Bell show?}\label{sec6.2}

 Bell's result is rather well known but often  misunderstood.
Bell showed that if we go back to EPR's dilemma:

 \begin{itemize}
	\item [1.] Either there exist ``hidden variables", namely pre-determined results of experiments (variables not included in the quantum state), 

\item [2.] Or there exists actions at a distance in Nature.
 \end{itemize}

    Then,  the first assumption alone leads to a contradiction with quantum predictions (very well verified by now) when different spin components are measured on the particle on Alice's side and the one on Bob's side\footnote{Actual experiments are done with photons and their polarizations, but both situations are conceptually similar.}. For a simple proof, see D\"urr et al. \cite{DGTZ}.

So, only the second possibility remains: there exists instantaneous actions at a distance in Nature.
 That is certainly the most ``crazy" of the two possibilities, but the only one that is tenable, given EPR's dilemma,  Bell's argument and the empirical verifications of the violation of his inequalities.

Note that there is no assumption in Bell's argument about:
   
  \begin{itemize}
	\item [1.] The existence of hidden variables,  
    
  \item [2.] Realism,

 \item [3.] Or determinism.

      \end{itemize}
   
 Indeed, the ``the existence of hidden variables" is deduced through the EPR dilemma from the assumption of locality and the last two expression, whenever they are precisely defined, are similar to the assumption of the existence of hidden variables.

Actually, Bell stated very clearly what he showed (here ``impasse" refers to nonlocality):

\begin{quote}

Let me summarize once again the logic that leads to the impasse. The EPRB correlations are such that the result of the experiment on one side immediately foretells that on the other, whenever the analyzers happen to be parallel.

    If we do not accept the intervention on one side as a causal influence on the other, we seem obliged to admit that the results on both sides are determined in advance anyway, independently of the intervention on the other side, by signals from the source and by the local magnet setting.

    But this has implications for non-parallel settings which conflict with those of quantum mechanics. So we {\it cannot\/} dismiss intervention on one side as a  causal influence on the other.

\begin{flushright}  John Bell  \cite[pp. 149--150]{B} (italics in the original)\end{flushright} 
\end{quote}

 \subsection{Misunderstandings of Bell}\label{sec6.3}
 
  However, Bell's result has been almost universally taken as a no hidden variable result, hence a proof that Bohr was right (against EPR) and that quantum mechanics is ``complete"\footnote{Actually, Bohr's answer to the EPR paper \cite{Bo3} is very unclear. For a critical analysis of that paper, see Bell \cite[p. 155-156]{B}.}.

In a narrow sense, this is true, if one forgets EPR's dilemma, which of course Bell never forgot. If one does not forget this  dilemma, Bell's result is a non-locality result (namely proving the existence of instantaneous actions at a distance) rather than a no hidden variable one.

 \subsection{Schr\"odinger's generalization of the EPR argument}\label{sec6.4}

Bell's argument relied on the statistics of measurements of the spin, or the polarization, in different directions for the particle on Alice's side and on Bob's side.
But those statistics had to be checked. Hence, last year's Nobel prizes.

Could one prove non-locality using only the perfect correlations?
 Yes, using Schr\"odinger's ideas in his 1935  paper \cite{Sch}\footnote{For more details about the reasonings below, see \cite{BGH1} or Hemmick \cite{Hem, Hem-S}; see also \cite{BGH2} for an extension of this result using the original EPR variables, positions and momenta, and see \cite{BGH3} for a pedagogical summary.}.
The main innovation of Schr\"odinger was to extend the previous EPR-type argument  to {\it all} observables, at least given suitable quantum states namely:

 {\bf Maximally entangled states}

Consider a finite dimensional (complex) Hilbert space $\cal H$, of dimension $N$.

A unit vector  $\Psi$ in $\cal H \otimes \cal H$ is {\it maximally entangled} if it is of the form:

 \be
\Psi= 
	\frac{1}{\sqrt N}
	\sum_{n=1}^N    \psi_n \otimes \phi_n,
\label{7}
\ee
where  $\psi_n$ and $\phi_n$ are orthonormal bases in $\cal H$.

Since we are interested in quantum mechanics, we will refer to those vectors as {\it maximally entangled states} and we will associate, by convention, each space in the tensor product to a ``physical system," namely we will consider the set  $\{\psi_n\}_{n=1}^N$ as a basis of states for physical system 1 (associated with Alice when measurements are made on that system) and the set  $\{\phi_n\}_{n=1}^N$ as a basis of states for physical system 2 (associated with Bob when measurements are made on that system).

Now, given a  maximally entangled state, one can  associate to each operator of the form $  O \otimes  {\bf  1}  $
 (meaning that it acts non-trivially only on particle 1, since here ${\bf  1}$ denotes the identity operator on  $\cal H$) an operator of the form ${\bf  1}  \otimes \tilde O$  (meaning that it acts non-trivially only on particle 2), so that:

If $\psi_n$ are eigenstates of $O$, with eigenvalues $\lambda_n$,
\be
 O \psi_n = \lambda_n \psi_n,
\label{8}
\ee
then, the states $\phi_n$ are eigenstates of $\tilde O$, also with eigenvalues $\lambda_n$:
\be
\tilde O \phi_n = \lambda_n \phi_n.
\label{9}
\ee
For the proof of this statement, see \cite{BGH1}.

 A simple example of a maximally entangled state is the one introduced in our discussion of EPR (\ref{6a}) (this shows that this notion of maximally entangled state is a proper extension of the state (\ref{6a})).

 In the previous notation, (\ref{7}), 
 one has:
 $$\psi_1= | \uparrow>,
$$
$$ \psi_2=-| \downarrow>.
$$
$$
\phi_1= |  \downarrow>,
$$
$$ \phi_2 = | \uparrow>. 
$$

If one takes 
\ba
O = \left(\begin{array}{ccc} 1 & 0 \\ 0 & -1 \end{array}\right)
\label{10}
\ea
which corresponds to the spin operator for system $1$ and  has eigenvectors $\psi_1$ with eigenvalue $1$ and $\psi_2$ with eigenvalue $-1$, it turns out that the operator $\tilde O$ is of the form:
\ba
\tilde O =  \left(\begin{array}{ccc} -1 & 0 \\ 0 & 1 \end{array}\right)=-O,
\label{11}
\ea
which means that the spin operators for systems $1$ and $2$ are perfectly anti-correlated, since $\tilde O$ is minus the spin operator for system $2$; this  of course was the situation considered by EPR.
 
Let us now generalize the EPR dilemma  by applying the previous result to spatially separated physical systems.
Suppose that we have a pair of physical systems, whose states belong to the same finite dimensional  Hilbert space $\cal H$. And suppose that the quantum state $\Psi$ of the pair is maximally entangled.

So, the quantum state $\Psi$ of the pair is of the form \ref{7}. 
Any ``observable" acting on system 1 is represented by a self-adjoint operator $O$, which has therefore a basis of eigenvectors.
It turns out that we may choose, without loss of generality,  the set $\{\psi_n\}_{n=1}^N$  as being   the eigenstates of $O$. Let  $\lambda_n$  be the corresponding  eigenvalues. 
 
 If one measures the corresponding observable $\tilde O$ on system $2$, the result will be one of the eigenvalues $\la_n$, each having equal probability $\frac{1}{N}$. If the result is $\la_k$,  the (collapsed) state of the system after the measurement, will be $\psi_k \otimes \phi_k $.
  
 Then, the measurement of observable $ O$, on system $1$, will necessarily yield the value $\la_k$.

So, we have again a situation (like for EPR) of:
 
{\bf Perfect correlations.}

In any maximally entangled  quantum state, there is, for each operator $O$ acting on system 1, an operator $ \tilde O$ acting on system 2, such that, if one measures the physical quantity represented by operator $ \tilde O$ on system 2 and the result is the eigenvalue  $\la_k$ of $ \tilde O$, then, measuring the physical quantity represented by operator 
$O$  on system 1 will yield with certainty the same eigenvalue $\la_k$ (and vice versa).

 Let's go back to EPR's dilemma. In order to explain those perfect correlations, we have only two possibilities:

\begin{itemize}
	\item [1.] Either the results on both sides are pre-determined before any measurement (there exist ``hidden variables") and are perfectly correlated.

\item [2.] Or a measurement on one side affects instantaneously (or creates) the result on the other side: actions at a distance!
 \end{itemize}

Let us  call this
 {\bf Schr\"odinger's dilemma}. 
 And let us make the first possibility more explicit:
 
For any operator $O$ acting on system $1$, there is a value $v(O)$ that a measurement of $O$ would reveal and not create, since it must pre-exist to any measurement. And such a pre-existing value $v(O)$ must exist {\it for every} $O$ acting on system $1$.

To summarize, we have shown:

\begin{theorem}\label{t2} (Schr\"odinger's Theorem.)

 Let $\cal A$ be the set of self-
adjoint operators on the component Hilbert space   $\cal H$ of a physical system 
in a maximally entangled state $\Psi$. Then, assuming locality and  the perfect correlations predicted by quantum mechanics, there 
exists a  value-map  $v: \cal A \to \R$, i.e. a map that assigns a value $v(O)$ to  any experiment associated with what is called in quantum mechanics a ``measurement of an observable O".

\end{theorem}
 
 The conclusion of Schr\"odinger's theorem \ref{t2} and of theorem \ref{t1} on the non-existence of  ``hidden variables" or of value-maps plainly contradict each other\footnote{The reason we need this general discussion about maximally entangled states, is that the theorem \ref{t1} on the non-existence of  ``hidden variables" holds only in spaces of  $4$ dimensions or more (actually $3$ dimensions or more would suffice if we use only assumption (\ref{4}), see Kochen and Specker \cite{KS}). So it would not be sufficient to consider only the maximally entangled state (\ref{6a}), since there the Hilbert space denoted $\cal H$  is two-dimensional.}.
So, the assumptions of at least one of them must be false. 
To derive Schr\"odinger's theorem, we assume only the perfect correlations and locality. The theorem on the non-existence of  value-maps is purely mathematical.

The  perfect correlations are an immediate consequence of quantum mechanics.
The only remaining assumption  is locality. Hence we can deduce:

 \begin{theorem}\label{t3} (Non-locality ``Theorem")

The locality assumption is false.

 \end{theorem}

Let us summarize the argument:

Given a maximally entangled state, for every operator acting on system $1$, there is a corresponding operator acting on system $2$ such that the results of the measurements of both operators are perfectly correlated. There are only two possibilities:

-The results of those measurements are
 pre-determined before any measurement.

-Or non-locality (instantaneous actions at a distance).

The first possibility implies the existence of a  value-map.
But this is impossible because of the Bell-Kochen-Specker no hidden variables theorem \ref{t1}.
The only remaining possibility  is non-locality.
    
Of course this is not a purely mathematical theorem, since it depends on a physical assumption: the perfect correlations associated to maximally entangled state.
  
Nevertheless, the only things to check experimentally are those perfect correlations. No need for the Clauser-Aspect-Zeilinger experiments on correlations between ``measurements" of spins or polarizations in different directions for Alice and Bob.

How to reconcile this non-locality with relativity is a difficult open question, see Maudlin \cite{Ma} for a detailed discussion.
    
 Note finally that a famous 19th century British scientist expressed the EPR-Bell-Schr\"odinger proof of nonlocality quite succinctly:

  \begin{quote}   
   When you have eliminated the impossible, whatever remains, however improbable, must be the truth.
    
\begin{flushright}   Sherlock Holmes \end{flushright}
\end{quote}

\section{Conclusions }\label{sec7}
The de Broglie-Bohm theory is not yet another ``interpretation" of quantum mechanics.
 It is not a different theory than quantum mechanics.
 It is simply the (rational) completion of quantum mechanics, which is manifestly incomplete since it does not speak of what happens outside of laboratories.

And ordinary quantum mechanics is simply de Broglie-Bohm's theory applied to what happens in laboratories.
Once this is understood, there is no need to run around in search of ``alternatives" to quantum mechanics or of ``interpretations" of it.

\end{document}